\pdfoutput=1

\documentclass[11pt,twoside,a4paper,cmspaper,final,collab]{cms-tdr}
\def\svnVersion{290987}\def\cmsCernNo{2013-037}\def\cmsCernDate{\today}\def\cmsMessage{Published in the European Physical Journal C as \href{http://dx.doi.org/10.1140/epjc/s10052-015-3435-4}{\doi{10.1140/epjc/s10052-015-3435-4.}}}

\begin{document}\cmsNoteHeader{HIN-12-017}

\hyphenation{had-ron-i-za-tion}
\hyphenation{cal-or-i-me-ter}
\hyphenation{de-vices}
\RCS$Revision: 283965 $
\RCS$HeadURL: svn+ssh://svn.cern.ch/reps/tdr2/papers/HIN-12-017/trunk/HIN-12-017.tex $
\RCS$Id: HIN-12-017.tex 283965 2015-04-09 18:14:42Z alverson $
\newlength\cmsFigWidth
\ifthenelse{\boolean{cms@external}}{\setlength\cmsFigWidth{0.98\columnwidth}}{\setlength\cmsFigWidth{0.6\textwidth}}
\ifthenelse{\boolean{cms@external}}{\providecommand{\cmsLeft}{top}}{\providecommand{\cmsLeft}{left}}
\ifthenelse{\boolean{cms@external}}{\providecommand{\cmsRight}{bottom}}{\providecommand{\cmsRight}{right}}
\newcommand{\etacm}{\ensuremath{\eta_\textsc{cm}\xspace}}
\providecommand{\rtsnn}{\ensuremath{\sqrt{s_{_\mathrm{NN}}}}\xspace}
\providecommand{\mbinv} {\ensuremath{\,\mathrm{mb}^{-1}}\xspace}
\cmsNoteHeader{HIN-12-017}
\title{Nuclear effects on the transverse momentum spectra of charged particles in pPb collisions at $\rtsnn=5.02$\TeV }

\date{\today}

\abstract{Transverse momentum spectra of charged particles are measured by the CMS experiment
at the CERN LHC in
pPb collisions at $\rtsnn=5.02$\TeV, in the range $0.4 < \pt < 120$\GeVc and pseudorapidity $\abs{\etacm}  < 1.8$ in the proton-nucleon center-of-mass frame. For  $\pt <10$\GeVc, the charged-particle production is asymmetric
about $\etacm = 0$, with smaller yield observed  in the direction of the proton beam,
qualitatively consistent with expectations from shadowing in nuclear parton distribution functions (nPDF).
A pp reference spectrum  at $\sqrt{s}=5.02$\TeV is obtained by interpolation from previous measurements at higher and lower center-of-mass energies. The  \pt distribution measured in pPb collisions shows an enhancement of charged particles
with  $\pt>20$\GeVc compared to expectations from the pp reference. The enhancement is larger
than predicted by perturbative quantum chromodynamics calculations that include antishadowing modifications of nPDFs.
}

\hypersetup{%
pdfauthor={CMS Collaboration},%
pdftitle={Nuclear effects on the transverse momentum spectra of charged particles in pPb collisions at a nucleon-nucleon center-of-mass energy of 5.02 TeV},%
pdfsubject={CMS},%
pdfkeywords={CMS, physics, relativistic heavy-ion collisions}}

\maketitle

\section{Introduction}
\label{sec:introduction}

The central goal of the heavy ion experimental
program at ultra-relativistic
energies is to create a system of
deconfined  quarks and gluons, known as a quark-gluon plasma (QGP), and to study its
properties as it cools down and transitions into a hadron gas. A key tool in the studies of the QGP is the phenomenon of jet quenching~\cite{Bjorken:1982tu}, in which the
partons produced in hard scatterings lose energy through gluon radiation and elastic scattering  in the hot and dense partonic medium~\cite{2010LanB...23..471D}. Since high transverse momentum quarks and gluons fragment into jets of hadrons, one of the observable
consequences of parton energy loss is the suppression of the yield of high-\pt particles
in comparison to their production in proton-proton (pp) collisions. This suppression, studied as a function of
the \pt and pseudorapidity ($\eta$) of the produced particle,
is usually quantified in terms of the nuclear modification factor,
defined as
\begin{equation}
R_\mathrm{AB}(\pt,\eta) = \frac{1}{\langle \mathrm{T}_\mathrm{AB} \rangle}
\frac{\rd^2 N^\mathrm{AB} / \rd\pt\, \rd\eta }{\rd^2 \sigma^{\Pp\Pp} / \rd\pt\, \rd\eta },
\label{eq:R_AB}
\end{equation}
where $N^\mathrm{AB}$ is the particle yield in a
collision between nuclear species A and B,
$\sigma^{\Pp\Pp} $ is the corresponding cross section in pp collisions, and $\langle \mathrm{T}_\mathrm{AB} \rangle$ is the
average nuclear overlap function~\cite{Miller} in the AB collision (in the case of proton-nucleus collisions, the quantity $\langle \mathrm{T}_\mathrm{AB} \rangle = \langle \mathrm{T}_\mathrm{pA} \rangle$ is called average nuclear thickness function).
If  nuclear collisions behave as incoherent superpositions of nucleon-nucleon collisions, a ratio of unity is expected. Departures from unity are
indicative of final-state effects such as parton energy loss, and/or initial-state effects such as
modifications of the nuclear parton distribution functions (nPDF)~\cite{Eskola:2009uj}.
The nPDFs are constrained by measurements in
lepton-nucleus deep-inelastic scattering (DIS)
and Drell--Yan (DY) production of dilepton pairs from $\PQq\PAQq$ annihilation in proton-nucleus collisions~\cite{Arneodo1994301}.
In the small parton fractional momentum regime ($x \lesssim  0.01$), the nPDFs are found to be suppressed
relative to the proton PDFs, a phenomenon commonly referred to as ``shadowing''~\cite{Frankfurt:1988nt}.
At small $x$, where the parton distributions are described theoretically by non-linear evolution equations in $x$, gluon saturation is predicted by the color glass condensate models~\cite{Iancu:2001md,Venugopalan:2003,Albacete:2013tpa}. For the $x$ regime $0.02 \lesssim x \lesssim 0.2$, the nPDFs are enhanced
(``antishadowing'') relative to the free-nucleon PDFs~\cite{Arneodo1994301}.

To gain access to the properties of the QGP produced in heavy ion
collisions it is necessary to separate the effects directly related to the
hot partonic medium from those that are not, referred to as ``cold nuclear matter''
effects.  In particular, nPDF effects are expected to play an important
role in the interpretation of nuclear modification factors at the CERN LHC.
Unfortunately, the existing nuclear DIS and DY measurements constrain only poorly the gluon
distributions over much of the kinematic range of interest.
High-\pt hadron production in
 proton-nucleus
(or  deuteron-nucleus) collisions provides a valuable reference for
nucleus-nucleus collisions, as it probes initial-state nPDF
modifications over a wide kinematic range and is expected to be
largely free from the final-state effects that accompany QGP
production~\cite{Salgado:2011wc}.

The measurements of the nuclear modification factors of neutral pions
and charged hadrons in the most central gold-gold (AuAu) collisions at the Relativistic
Heavy Ion  Collider (RHIC)~\cite{Adler:2003qi,Adler:2003au,Back:2003qr,Adams:2003kv} revealed a large suppression at high
\pt, reaching an $R_\mathrm{AA}$ as low as 0.2. In contrast, no such suppression was
found at mid-rapidity in deuteron-gold collisions at the same
energy~\cite{BRAHMS:2003,PHENIX:2003,PHOBOS:2003,STAR:2003a}. These findings established parton energy loss,
rather than initial-state effects~\cite{Kharzeev:2003},
 as the
mechanism responsible for the modifications observed in AuAu collisions.

{\tolerance=600
At the LHC, the
charged-particle suppression in lead-lead (PbPb) collisions
persists at least up to a \pt of
100\GeVc~\cite{CMS:2012,ALICE:2013}. In proton-lead (pPb) collisions, the ALICE Collaboration reported
no significant deviations from unity in the charged-particle $R_\mathrm{pPb}$ up to $\pt\approx50$\GeVc~\cite{Abelev:2014dsa}.
The analysis presented here used data from CMS to extend the measurement of the
charged-particle $R_\mathrm{pPb}$ out to $\pt \approx 120$\GeVc, with
the aim of evaluating initial-state effects over
a kinematic range similar to that explored through measurements in PbPb collisions~\cite{CMS:2012}.
\par}

Proton-nucleus collisions have already been used to assess the impact of cold nuclear matter on
jet production at the LHC. The transverse momentum balance, azimuthal angle correlations,
and pseudorapidity distributions of dijets have been measured as a function of the event activity,
and no significant indication of jet quenching was found~\cite{Chatrchyan:2014hqa}.
When normalized to unity, the minimum-bias dijet pseudorapidity distributions
are found to be consistent with next-to-leading-order (NLO) perturbative quantum chromodynamic (pQCD)
calculations only if  nPDF modifications are included~\cite{helenius2012impact}.
Similarly, inclusive jet $R_\mathrm{pPb}$ measurements are also found to be consistent
with NLO pQCD predictions that include nPDF modifications~\cite{Aad:2014bxa,Adam:2015hoa}.
The measurement of the charged-particle spectra presented in this paper provides a comparison
to theory that is sensitive to smaller $x$ values than those accessible in the jet
measurements. However, it should be noted that the charged-particle $R_\mathrm{pPb}$ is
dependent upon non-perturbative hadronization effects, some of which, such as
gluon fragmentation into charged hadrons, are poorly constrained  at the LHC energies~\cite{d'Enterria:2013vba}.

{\tolerance=800
The \pt distributions of inclusive charged particles in pPb collisions at a
nucleon-nucleon center-of-mass
energy of 5.02\TeV
are presented in the range of
$0.4<\pt<120$\GeVc. The measurement is performed in several pseudorapidity intervals over $\abs{\etacm}< 1.8$.
Here $\etacm$ is the pseudorapidity in the proton-nucleon center-of-mass frame.
The nuclear modification factor is studied at mid-rapidity, $\abs{\etacm}< 1$,
and the forward-backward asymmetry of the yields, $Y_\text{asym}$, defined as
\begin{equation}
Y_\text{asym}^{(a,b)}(\pt) = \frac{\int_{-b}^{-a}\rd\etacm \, \rd^2 N_\text{ch}(\pt) / \rd\etacm\, \rd\pt}
                    {\int_{a}^{b}\rd\etacm \, \rd^2 N_\text{ch}(\pt) / \rd\etacm\, \rd \pt},
\label{eq:Y_asym}
\end{equation}
is presented for three pseudorapidity intervals,
where $a$ and $b$ are positive numbers,
and $ N_\text{ch}$ is the yield of charged particles.
\par}

Due to their wide kinematic coverage, the measurements are expected to  provide information about the
nPDFs in both the shadowing and antishadowing regions. In particular, the effects of shadowing are expected to be more prominent
at forward pseudorapidities (in the proton-going direction), where smaller $x$ fractions in the nucleus are accessed.

In the absence of other competing effects, shadowing in the Pb nPDFs would
result in values of $Y_\text{asym}>1$ at low \pt (\ie, small $x$).
The effects of antishadowing can be probed with the  $R_\mathrm{pPb}$ measurement
at larger \pt values of $30 \lesssim  \pt \lesssim 100$\GeVc that correspond
approximately to $ 0.02 \lesssim x \lesssim 0.2$. Antishadowing in the nPDFs may increase the yield of
charged particles in pPb collisions compared with  expectations from the yield in pp collisions.

\section{Data selection and analysis}
\label{sec:analysis}

\subsection{Experimental setup}
\label{sec:experiment}

A detailed description of the CMS detector can be found in Ref.~\cite{JINST}.
The CMS experiment uses a right-handed coordinate system,
with the origin at the nominal interaction point (IP) at the center of the detector, and the $z$ axis along the beam direction.
The silicon tracker, located within the 3.8\unit{T}
magnetic field of the superconducting solenoid, is used to reconstruct charged-particle tracks.
Consisting of 1440 silicon pixel detector modules and 15\,148 silicon
strip detector modules, totaling about 10 million silicon strips and 60 million
pixels, the silicon tracker measures the tracks of charged
particles within the pseudorapidity range $\abs{\eta}< 2.5$. It provides
an impact parameter resolution of $\approx 15\mum$ and a
\pt resolution of about 1.5\% for particles with \pt of 100\GeVc.
An electromagnetic calorimeter (ECAL) and a hadron calorimeter (HCAL) are also
located inside the solenoid. The ECAL consists of more than
75\,000 lead tungstate crystals, arranged in a quasi-projective
geometry; the crystals are distributed in a barrel region ($\abs{\eta} < 1.48$) and in
two endcaps that extend out to $\abs{\eta} \approx 3.0$. The HCAL barrel and
endcaps, hadron sampling calorimeters composed of brass and scintillator
plates, have an acceptance of $\abs{\eta} \lesssim 3.0$. The hadron forward calorimeters (HF),
consisting of iron with quartz fibers read out by photomultipliers, extend the calorimeter
coverage out to $\abs{\eta} = 5.2$, and are used in offline
event selection. Beam Pick-up Timing for the eXperiments (BPTX) devices
were used to trigger the detector readout. They are located around the beam pipe at a distance
of 175\unit{m} from the IP on either side, and are designed to provide precise information on the
LHC bunch structure and timing of the incoming beams.
The detailed Monte Carlo (MC) simulation of the CMS detector response is based on
\GEANTfour~\cite{bib_geant}.

This measurement is based on a data sample corresponding to an
integrated luminosity of 35\nbinv, collected
by the CMS experiment in pPb collisions during the 2013 LHC running period.
The center-of-mass energy per nucleon pair was
$\rtsnn=5.02$\TeV, corresponding to per-nucleon beam energies of 4\TeV and
1.58\TeV for protons and lead nuclei, respectively.
The data were taken with two beam configurations. Initially, the
Pb nuclei traveled in the counterclockwise direction, while in the
second data-taking period, the beam directions were reversed. Both
data sets, the second one being larger by approximately 50\%, were analyzed independently, yielding
compatible results.
To combine data from the two beam configurations,
results from the first data-taking period are reflected along the $z$-axis,
so that in the combined analysis, the proton travels in the positive $z$ and $\eta$
directions. In this convention, massless particles emitted at $\etacm = 0$
in the nucleon-nucleon center-of-mass frame will be detected at
$\eta_\text{lab} = 0.465$ in the laboratory frame. A symmetric region about $\etacm = 0$ is used in the data analysis, resulting in a selected pseudorapidity range of $\abs{\etacm} < 1.8$.

\subsection{Event selection}

The CMS online event selection employs a hardware-based level-1 (L1) trigger and a software-based high-level
trigger (HLT). A minimum-bias sample is selected first by the L1 requirement of a pPb bunch
crossing at the IP (as measured by the BPTX), and an HLT requirement of at least one reconstructed track with $\pt>0.4$\GeVc in the
pixel tracker. For most of the 5.02\TeV data collection, the minimum-bias trigger is significantly
prescaled because of the high instantaneous LHC luminosity. To increase the
\pt reach of the measurement, a set of more selective triggers is also used: additional L1 requirements are imposed to select events that have at
least one reconstructed calorimeter jet with an uncorrected transverse energy of $\ET > 12$\GeV,
and $\ET > 16$\GeV. These event selections are complemented by additional HLT requirements that select events
based on the presence of at least one track with $\pt > 12$\GeVc (for L1 $\ET > 12\GeV$),
or with $\pt > 20$ or
30\GeVc (for L1 $\ET > 16\GeV$) reconstructed in the pixel and strip tracker.

The above triggering strategy allows for the optimization of the data-taking rate while adequately sampling all \pt regions,
including collecting all events containing  very high-\pt tracks. The track trigger with a \pt threshold of 12\GeVc records
about 140 times more events with high-\pt tracks than the minimum-bias trigger, the track $\pt>20$\GeVc trigger enhances
this with an additional factor of about 8, while the
track $\pt>30$\GeVc trigger that is not prescaled, increases the
number of events with a high-\pt track by yet another factor of about 2.

In the offline analysis, additional requirements are applied.
Events are accepted if they have (i) at least one HF calorimeter tower on both the positive and
negative sides of the HF with more than $3\GeV$ of total energy, (ii) at least one reconstructed
collision vertex with two or more associated tracks, and (iii) a maximum distance of 15\cm along the beam axis
between the vertex with the largest number of associated tracks and the nominal IP. Beam-related
background is suppressed
by rejecting events
where less than 25\% of all reconstructed tracks are of good quality~\cite{Khachatryan:2010pw}.

An event-by-event weight factor accounts for correcting the measured charged-particle spectra in pPb collisions to a detector-independent class of collisions termed as ``double-sided'' (DS) events,
which are very similar to those that pass the offline selection described above. A DS event is
defined as a collision producing at least one particle in the pseudorapidity range $-5<\eta_\text{lab}<-3$
and another
in the range $3<\eta_\text{lab}<5$,
each with proper lifetime $\tau > 10^{-18}$\unit{s} and energy $E > 3$\GeV~\cite{Chatrchyan:2013eya}.
The performance of
the minimum-bias and high-$\pt$ single-track triggers, as well as the offline criteria in selecting DS
events, is evaluated with simulations using the \HIJING MC
generator~\cite{Wang:1991hta}, version 1.383~\cite{Gyulassy:1994ew}, and the correction factors are computed as a function of the event
multiplicity. An efficiency of 99\% is obtained for
the minimum-bias trigger and a negligible correction (\ie, 100\% efficiency) for the high-\pt track-triggered
events.
The correction factor is also evaluated using an \textsc{epos}~\cite{Werner:2005jf} simulation and,
based on the difference between both generators, a slightly \pt-dependent  systematic uncertainty
of 1\%
is
assigned to the final spectra.

During the pPb data taking period, about 3\% of the recorded events contained more than one pPb collision.
To reduce potential bias in the measurements arising from such ``pileup'',
events with multiple reconstructed vertices are removed if the longitudinal distance between the vertices
along the beamline is greater than a specific value that is related to the uncertainty of the vertex position. This value
is also dependent on the number of tracks associated with each vertex and ranges from 0.2\cm for vertex
pairs with at least 25 tracks associated with each vertex, to 3\cm for vertex pairs with only 3 tracks
associated with the vertex having the fewest associated tracks.
Simulated \HIJING events are used to tune the pileup-rejection algorithm
in order to reduce the number of erroneously eliminated single-collision events
to a negligible fraction, and still maintain a high rejection efficiency
for genuine pileup events. The pileup-rejection efficiency is found to be $92\pm2\%$,
which is confirmed by using a low bunch intensity control sample in data.

\begin{figure}[t!h]
  \begin{center}
    \includegraphics[width=0.48\textwidth]{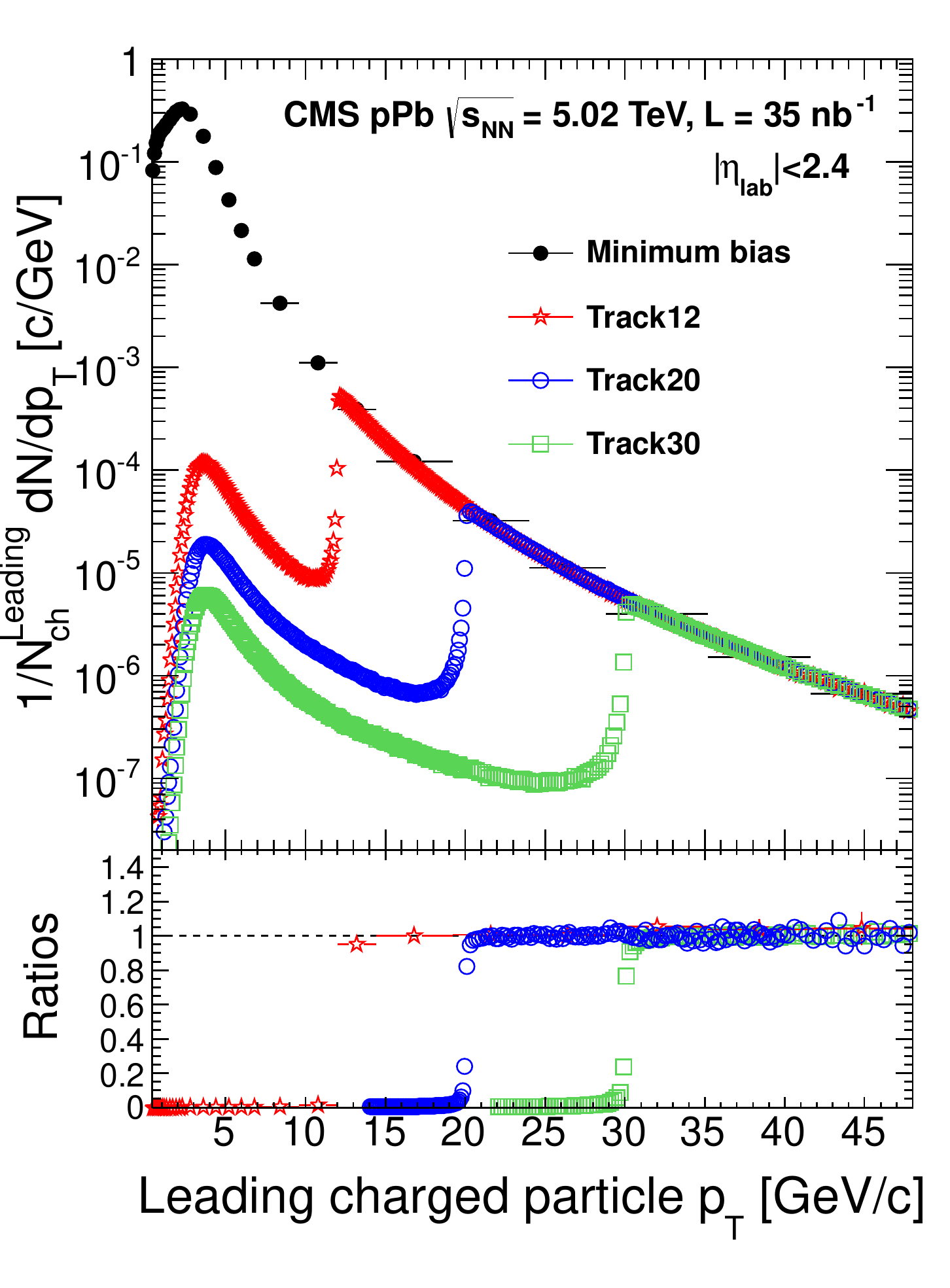}
    \caption{Top: Charged-particle yields for the different triggers normalized to the number of leading charged particles with $\pt>0.4$\GeVc in double-sided events, $N_\text{ch}^\text{Leading}$, as a function of leading-track \pt. The
      track-triggered distributions are normalized by the number of leading tracks in regions not affected by
      the rapid rise of the trigger efficiency near threshold.
      Bottom: Ratios of the leading-track \pt distributions for the four different triggers. The stars indicate the
      ratio of the 12\GeVc over the minimum-bias samples, the circles the 20 over the 12\GeVc samples, and the squares the ratio of the 30 over the 20\GeVc track-triggered spectra.}
    \label{fig:trackTriggersSpectra}
  \end{center}
\end{figure}

To obtain inclusive particle spectra up to $\pt\approx 120$\GeVc, data recorded
with the minimum-bias and high-\pt track triggers must be combined appropriately.
The corresponding weight factors are computed by counting the number of events that contain leading
tracks (defined as the track with the highest \pt in the event) in the range of $\abs{\eta_\text{lab}}<2.4$ with \pt
values in regions not affected by trigger thresholds, \ie, where the trigger efficiency of the
higher-threshold trigger is constant relative to that of the lower-threshold trigger. The ratio of the
number of such events in the two triggered sets of data
are used as weight factors. For example, the region
above which the track trigger with a \pt threshold of 12\GeVc has constant efficiency is determined by
comparing the \pt distribution of the leading tracks to that of the minimum-bias data. Similarly, the
constant efficiency region of the 20\GeVc track trigger is determined by comparison to the 12\GeVc
track trigger, and the 30\GeVc trigger to the 20\GeVc trigger. The regions of constant efficiency for
each trigger, as a function of leading charged-particle \pt, are shown in
Fig.~\ref{fig:trackTriggersSpectra}. The 12, 20, and 30\GeVc triggers have constant efficiencies above
a leading charged-particle \pt of 14, 22, and 32\GeVc, respectively. The weight factors are then
computed using the leading-track \pt classes of $14<\pt<22$\GeVc, $22<\pt<32$\GeVc,
and $\pt>32$\GeVc for the three high-\pt triggers. The combined uncertainty in these
normalizations is dominated by the matching of the 12\GeVc track-triggered events to the minimum-bias
events.

Some events selected by the track triggers in Fig.~\ref{fig:trackTriggersSpectra} are observed to result in a
leading charged-particle \pt below the corresponding trigger threshold. This can happen if
the $\eta$ of the track above threshold is outside the $\eta$ range considered in the analysis,
and because
the final track reconstruction---described in Section~\ref{sec:tracks}---is more robust and selective than the
fast-tracking algorithm implemented in the HLT. When the HLT selects an event based on a misreconstructed
track, it is often the case that the track is not found in the final reconstruction.
To determine the inclusive particle spectrum, events are first uniquely classified into leading-track \pt classes in the pseudorapidity range in which the spectrum is being measured. The spectra are
constructed by taking events from the minimum-bias, 12\GeVc track, 20\GeVc track, and
30\GeVc track trigger, respectively, for each bin. A 4\% systematic uncertainty on the possible trigger-bias effect is estimated from MC simulations.
This procedure was verified in a data-driven way by constructing a
charged-particle spectrum from
an alternative combination of event samples triggered by reconstructed jets.
Both final spectra,
triggered by tracks and jets, are found to be consistent within the
associated systematic
uncertainty.

\subsection{Track reconstruction}
\label{sec:tracks}

The \pt distribution in this analysis corresponds to that of primary charged particles, defined as
all charged particles with a mean proper lifetime greater than 1\cm/$c$, including the products of strong and
electromagnetic decays, but excluding particles originating from secondary interactions in the detector
material. Weak-decay products are considered primary charged particles only if they are the
daughters of a particle with a mean proper lifetime of less than 1\cm/$c$, produced in the collision.

Charged particles are reconstructed using the standard CMS combinatorial track finder~\cite{Chatrchyan:2014fea}.
The proportion of misreconstructed tracks in the sample is reduced by applying an optimized set of
standard tracking-quality selections, as described in Ref.~\cite{Chatrchyan:2014fea}.
A reconstructed track is considered as a primary charged-particle
candidate if the statistical significance of the observed distance of closest
approach between the track and the reconstructed collision vertex is less than
three standard deviations, under the hypothesis that the track originated from this vertex.
In case an event has multiple reconstructed collision vertices but is not
rejected by the pileup veto, the track distance is evaluated relative to the
best reconstructed collision vertex, defined as the one associated with the
largest number of tracks, or the one with the
lowest $\chi^{2}$ if multiple vertices have the same number of associated tracks.
To remove tracks with poor momentum reconstruction,
the relative uncertainty of the momentum measurement $\sigma(\pt)/\pt$ is required
to be less than 10\%. Only tracks that fall in the kinematic range of $\abs{\eta_\text{lab}}<2.4$
and $\pt > 0.4$\GeVc are selected for analysis to ensure high tracking efficiency (70--90\%) and low misreconstruction rates ($<$2\%).

The yields of charged particles in each \pt and $\eta$ bin are weighted by a
factor that accounts for the geometrical acceptance of the detector, the
efficiency of the reconstruction algorithm, the fraction of tracks
corresponding to a non-primary charged particle, the fraction of
misreconstructed tracks that do not correspond to any charged particle,
and the fraction of multiply-reconstructed tracks, which belong to the
same charged particle.

The various correction terms are estimated using simulated minimum-bias
pPb events from the \HIJING event generator.
To reduce the statistical uncertainty in the correction factors at high \pt, samples of \HIJING events are
also mixed with pp dijet events from the \PYTHIA MC generator~\cite{bib_pythia} (version 6.423, tune D6T with CTEQ6L1 PDF for 2.76\TeV, tune Z2 for 7\TeV~\cite{Field:2010bc}).

The efficiency of the charged-particle reconstruction as well as the misreconstruction rates are also
evaluated using pPb events simulated with \textsc{epos}. Differences between the two MC models are
mostly dominated by the fraction of charged particles consisting of strange and multi-strange baryons
that are too short-lived to be reconstructed unless they are produced at very high \pt. Such
differences in particle species composition, which are largest for particles with $3\lesssim \pt
\lesssim 14$\GeVc, are propagated as a systematic uncertainty in the measured spectra. Below this
\pt range, the strange baryons are only a
small fraction of the inclusive charged particles in either model, and the difference in reconstruction
efficiency between particle species has less impact at even larger \pt, as high-\pt multi-strange
baryons can be directly tracked with high efficiency. Additional checks were performed by changing
cutoffs imposed during track selection and in the determination of the corresponding MC-based
corrections.
The corresponding variations in the corrected yields amount to 1.2--4.0\%
depending on the \pt region under consideration,
and are included in the systematic uncertainty.

Finite bin-widths and finite transverse momentum resolution can deform a steeply falling \pt
spectrum. The data are corrected for the finite bin-width effect as they will be compared to a pp reference spectrum obtained by interpolation. The binning corrections
are derived by fitting the measured
distribution and using the resulting fit function as a probability distribution to
generate entries in a histogram with the same \pt binning as used in the measurement. The correction factors are then obtained from the ratio of entries in the
bins of the
histogram to the fit function evaluated at the centers of the bins. This correction amounts to 0--12\%, depending on \pt.
A similar method is used to evaluate the ``smearing'' effect of the finite \pt resolution on the binned
distributions. It is found that the momentum measurement, which has a
resolution of $\sigma(\pt)/\pt \approx 1.5\%$ near a \pt of 100\GeVc, is sufficiently precise to only have a negligible effect on the measured spectra
and therefore no correction factor is applied. To account for
possible incorrect determination of the momentum resolution from the simulation,
the effects were again evaluated after increasing the value of $\sigma(\pt)/\pt$ by an additional 0.01,
which produces a maximal distortion in the spectrum at a given \pt of less than 1\%.

\subsection{Proton-proton reference spectrum}

The pPb collisions occur at a center-of-mass energy of 5.02 TeV per nucleon pair. At
this collision energy, no proton-proton collisions have been provided by particle accelerators yet.
The pp results closest in
center-of-mass energy ($\sqrt{s}$) and with similar acceptance are those measured at 2.76 and 7\TeV by the CMS
experiment~\cite{CMS:2012,QCD_spectrum_09_7}. The determination of the nuclear modification factor $R_\mathrm{AB}$
resides in an interpolated reference spectrum to be constructed from data at higher and lower energies.
We follow the direct interpolation method developed in Ref.~\cite{QCD_spectrum_09_7} using measured \pt
spectra from inelastic collisions with $\abs{\eta}<1.0$ at $\sqrt{s}=0.63$, 1.8, and 1.96\TeV collision energies from CDF~\cite{cdf_630_1800,cdf_1960},
and 0.9, 2.76, and 7\TeV collision energies from CMS~\cite{QCD_spectrum_09_7,CMS:2012}. This interpolation can
be performed either as a function of \pt or as a function of $x_\mathrm{T}\equiv 2p_\mathrm{T}c/\sqrt{s}$.

\begin{figure*}[t!h]
  \begin{center}
    \includegraphics[width=0.48\textwidth]{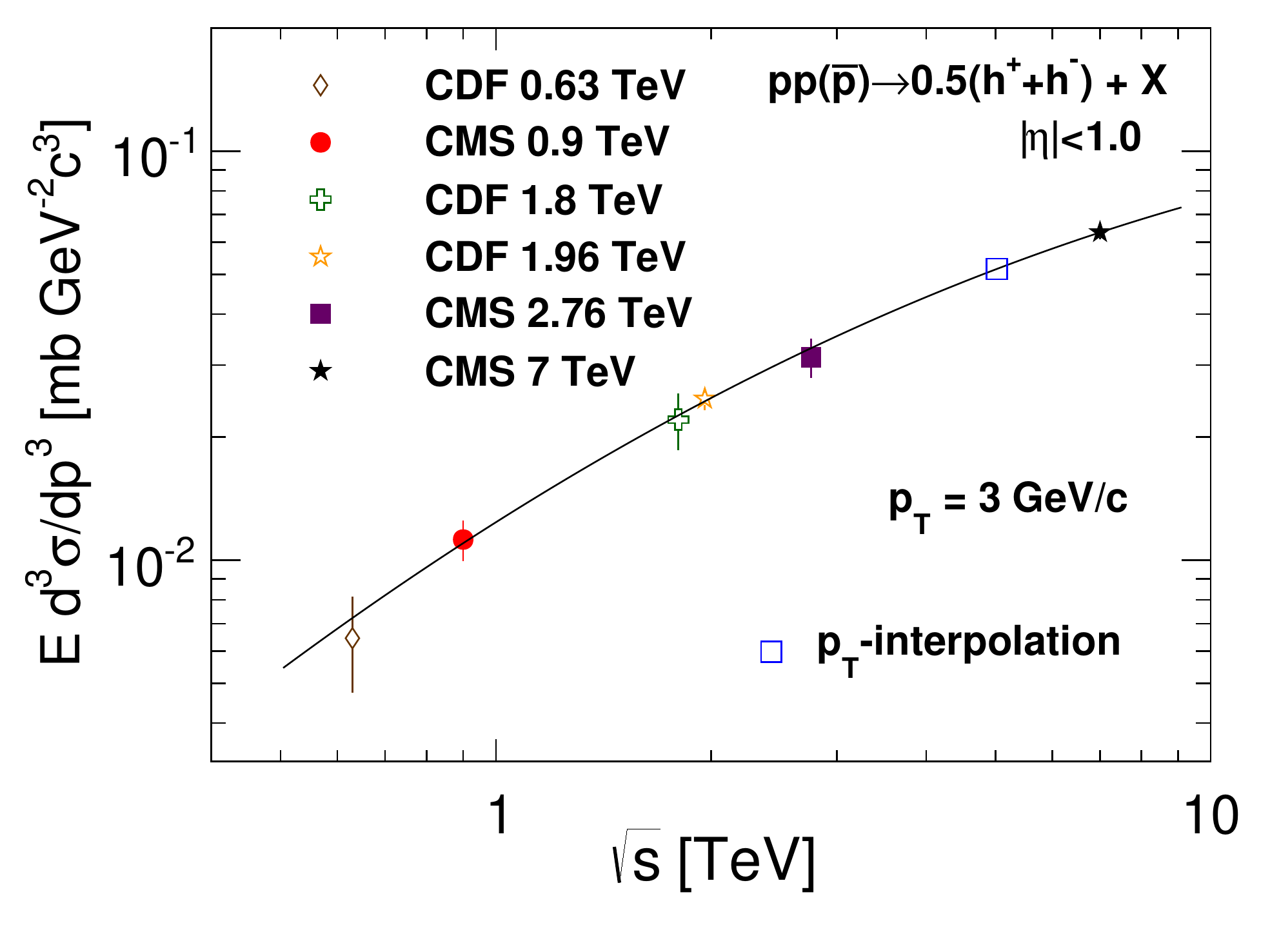}
    \includegraphics[width=0.48\textwidth]{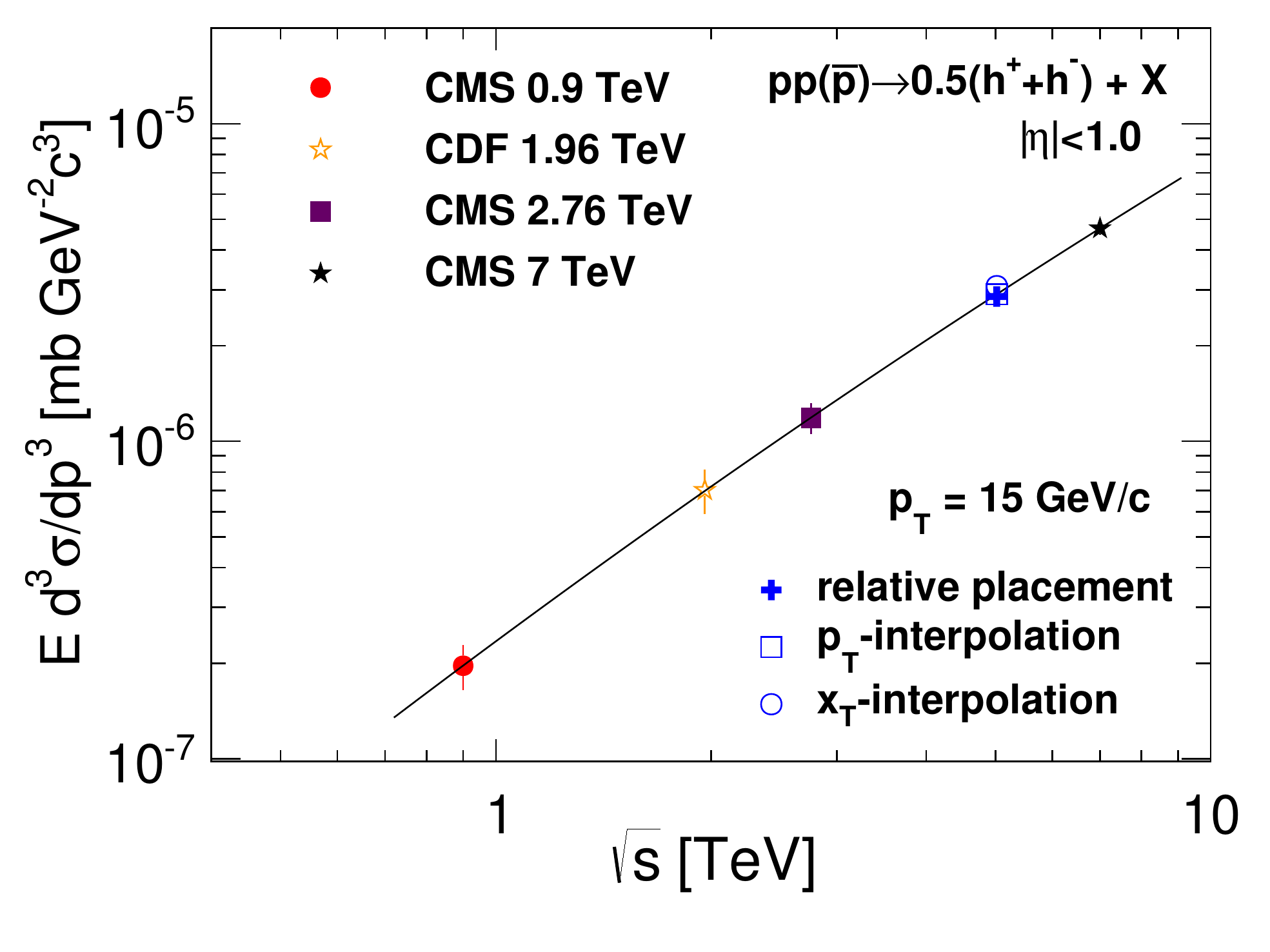}
    \includegraphics[width=0.48\textwidth]{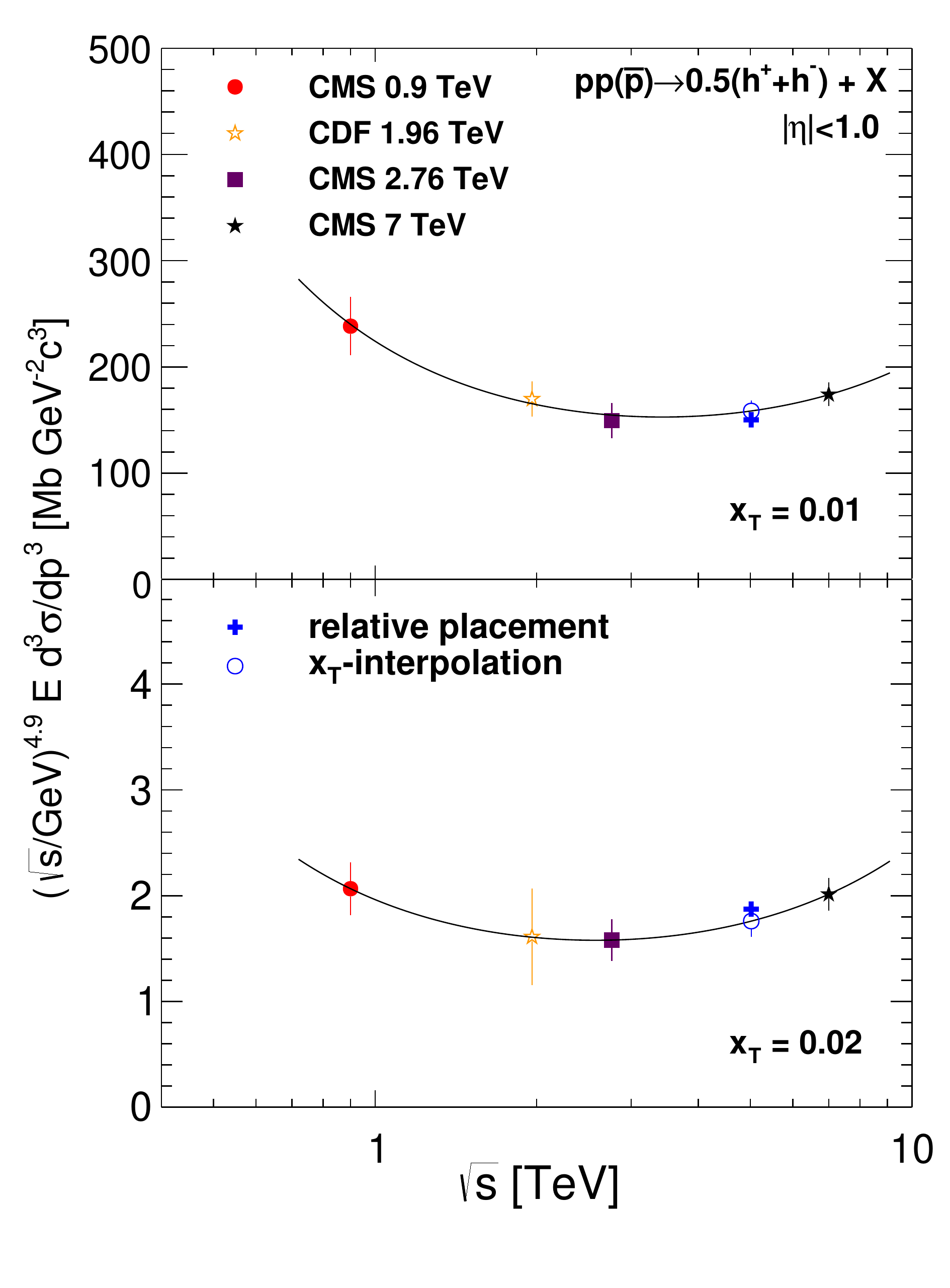}
    \includegraphics[width=0.48\textwidth]{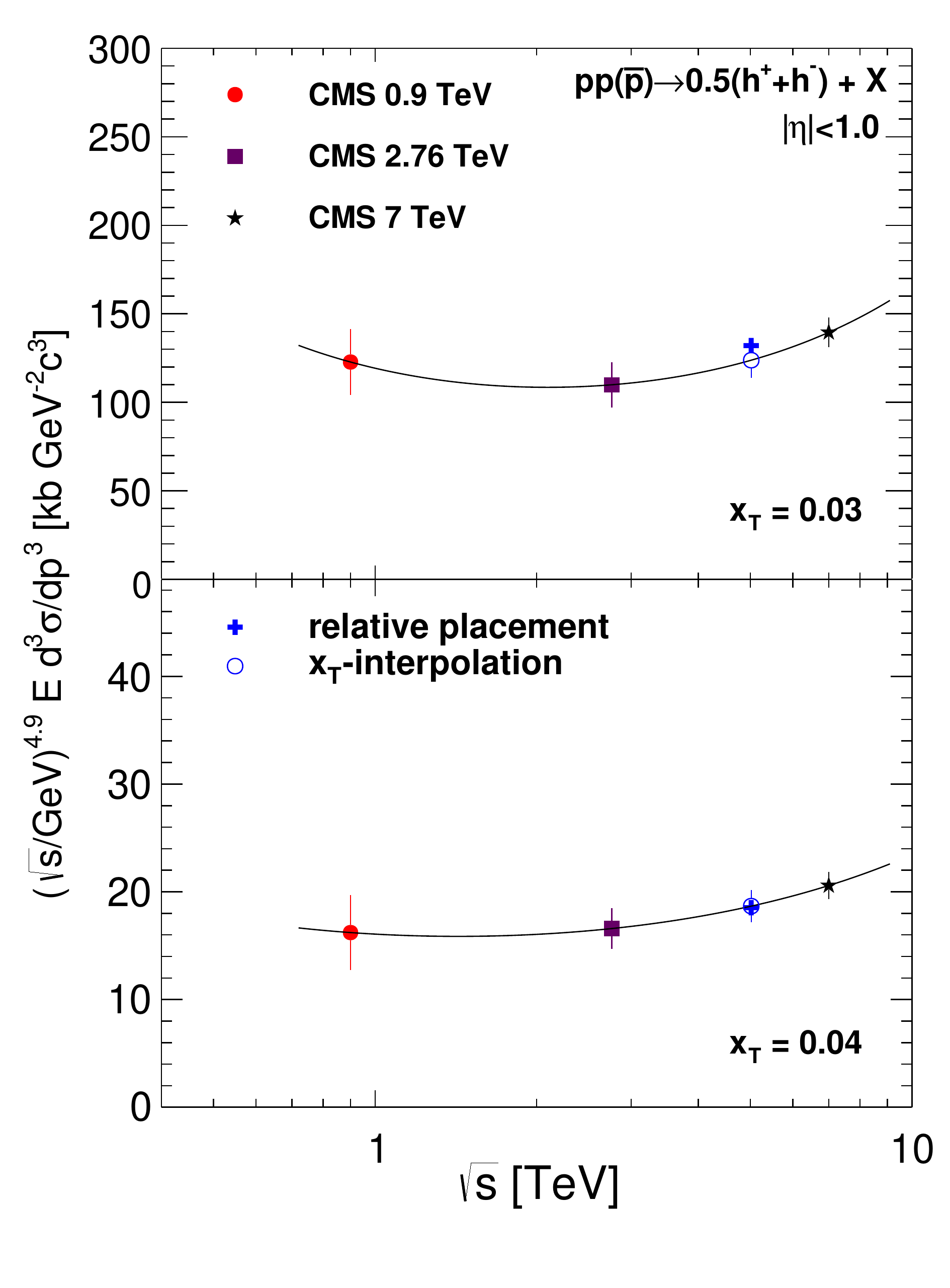}
    \caption{Examples of interpolations between measured charged-particle
differential cross sections at different $\sqrt{s}$ for \pt
values of 3 and 15\GeVc (top left and right), $x_\mathrm{T}$ values of 0.01
and 0.02 (bottom left), and $x_\mathrm{T}$ values of 0.03 and 0.04 (bottom right).
These $x_\mathrm{T}$ values correspond to $\pt \approx 25$, 50, 75, and 100\GeVc at $\sqrt{s}=5.02$\TeV.
The second-order polynomial fits, performed in the plane of the log-log invariant production vs.~$\sqrt{s}$, are shown by the solid
lines. The open squares and  circles, and the filled crosses
represent interpolated cross section values at 5.02\TeV using different
methods: \pt-based interpolation, $x_\mathrm{T}$-based interpolation, and
relative placement, respectively. The error bars on the interpolated
points represent the uncertainties in the fit.}
    \label{fig:ppRef}
  \end{center}
\end{figure*}

Since the \pt or $x_\mathrm{T}$ values of the input data points are often different for each measurement performed at the various collision energies, each spectrum must first be fitted as a function of \pt or $x_\mathrm{T}$.
An interpolation is performed by fitting each of the spectra to
a power-law dependence, and the resulting residuals to first- or third-order splines.
The fitted spectra are then interpolated to determine the value of the
reference spectrum at $\sqrt{s}=5.02$\TeV using a second-order
polynomial in the plane of the log-log invariant production \vs $\sqrt{s}$, as
shown in Fig.~\ref{fig:ppRef}. For the \pt-based direct
interpolation, data from only two of the six spectra are available at
$\pt > 30$\GeVc, which implies that the \pt-based direct interpolation is well
constrained only at low \pt. On the other hand, the $x_\mathrm{T}$-based
interpolation is well constrained at high \pt for $\sqrt{s}=5.02\TeV$,
so it is natural to combine the reference distributions from these
two direct interpolation methods.

The final pp reference spectrum is obtained by combining the \pt- and $x_\mathrm{T}$-based
reference spectra as follows. The \pt-based reference is chosen for \pt below 12.5\GeVc,
and the $x_\mathrm{T}$-based result above 13.5\GeVc; between these two \pt values a linear
weighting is implemented for the two references.
The systematic uncertainty in the pp reference spectrum is
determined through changing both the
specific method of interpolation, as well as the underlying pp reference data within their statistical and systematic
uncertainties.
The systematic uncertainty is dominated by the interpolation method, and is determined by
 comparing the combined \pt- and $x_\mathrm{T}$-based
reference spectra to the reference spectra obtained solely from the \pt
or $x_\mathrm{T}$ distributions, and also from a reference spectrum
determined by a ``relative placement'' method. In the latter, the
reference spectrum is obtained by computing where the 5.02\TeV spectrum
is situated with respect to the 2.76 and 7\TeV spectra in
\PYTHIA, and applying the computed placement factors to the
measured 2.76 and 7\TeV spectra. The placement factors are determined by
taking the value of the 5.02\TeV{} \PYTHIA spectrum, subtracting
the value of the 2.76\TeV spectrum, and dividing by the difference
between the 7 and the 2.76\TeV spectra. This process is then reversed by
using the computed placement factors from \PYTHIA, and replacing
the 2.76 and 7\TeV{} \PYTHIA spectra with the measured ones to
determine the interpolated 5.02\TeV spectrum.
Additionally, the NLO-based center-of-mass energy
rescaling proposed in Ref.~\cite{Arleo:2010kw} is found to yield results
consistent within the uncertainties of the other employed methods.
The uncertainty in the pp reference distribution due to the interpolation
method is estimated to amount to 10\%,
which captures the overall point-to-point variations in all of the interpolation
and scaling methods employed.
The contribution from the
uncertainties in the underlying pp input data corresponds to 6\%. These numbers
are added in quadrature, resulting in the 12\% uncertainty quoted for the
$\sqrt{s}=5.02$\TeV interpolated pp reference spectrum.

\section{Systematic uncertainties}

A summary of all the contributions to systematic uncertainties in the
\pt spectra, $R^{\ast}_\mathrm{pPb}$, and $Y_\text{asym}$ are given in
Table~\ref{tab:systematics}. The asterisk symbol is introduced to denote
that an interpolated, rather than measured, pp reference spectrum is used to
construct the nuclear modification factor.
Aside from the uncertainty from the spectra relative normalization and average
nuclear thickness, all uncertainties are determined by taking the approximate
maximum deviation from the central value found for the given source. For the particle species uncertainty, an
asymmetric uncertainty band is quoted because the maximum deviation above the
central value of the measurement is much larger than the maximum deviation
below. For the purpose of determining the significance of observed features in
the measurement,  the uncertainties are conservatively treated as following a
Gaussian distribution with a root mean square given by the value of the
uncertainty as determined above.

The degree of correlation among
different uncertainties
is described next.
For the spectra and $R^{\ast}_\mathrm{pPb}$ measurements, the uncertainty in the
efficiency of the single-track trigger and offline requirements in
selecting DS events is largely a normalization uncertainty, although it
also slightly affects the shape of the spectrum for $\pt \lesssim 3 \GeVc$. The
uncertainty from the contribution of the various
particle species to the unidentified spectrum
has the most significant effect in the region
$3 < \pt < 14 \GeVc$ and can impact the shape of the spectrum in a smooth fashion.
At high \pt, this effect is less prominent because, due to time dilation, unstable
particles have a higher probability of traversing the inner tracker before decaying and therefore a higher probability of being reconstructed.  Therefore, from this uncertainty the lower
bound on the pPb spectra measurement at higher \pt is 2.5\% below the central value,
which corresponds to no unstable particles being produced.
Uncertainty in track misreconstruction
can also affect the shape of the measured spectrum, as the misreconstructed
fraction of high-\pt particles
is sensitive to large occupancy in the silicon tracker within the cones of high-energy jets.
The uncertainty in tracking
selection can also affect the shape of the spectrum by
raising or lowering the measured values at high \pt, without changing the
low-\pt values, as high-\pt tracks are more sensitive to possible mismodeling of detector alignment than low-\pt tracks.
The uncertainty in the relative normalization of spectra is computed from the normalization factors involved in the combination of the \pt
distributions from different triggers. This uncertainty only applies for
selected \pt regions, and may raise or lower the spectrum above $\pt=14$\GeVc
by a constant factor of 1\% relative to the lower-\pt part of
the spectrum. The uncertainty from potential biases of the method used to combine triggers
can also affect the shape of the spectrum above \pt = 14\GeVc.

For the $R^{\ast}_\mathrm{pPb}$ measurement, the uncertainty in the average nuclear thickness function~\cite{Miller}
can influence the $R^{\ast}_\mathrm{pPb}$ curve by a constant multiplicative factor.
The uncertainty from the pp interpolation is strongly correlated among
points close together in \pt,
while some partial correlation remains throughout the whole \pt
region, even for very different \pt values.

For the forward-backward asymmetry measurements,
most of these uncertainties cancel in
part or in full when the ratio of the spectra is taken. The remaining uncertainty
in the detector acceptance and
tracking efficiency can change the shape of the forward-backward asymmetry, particularly at high \pt.

\begin{table}[t!h]\renewcommand{\arraystretch}{1.2}\addtolength{\tabcolsep}{-1pt}
\centering
\caption{Systematic uncertainties in the measurement of charged-particle spectra, $R^{\ast}_\mathrm{pPb}$, and $Y_\text{asym}$. The
ranges quoted refer to the variations of the uncertainties as a function of \pt.
Values in parentheses denote the negative part of the asymmetric uncertainty where applicable.
The total uncertainties of the measured pPb and the
interpolated pp spectra, as a function of \pt, are shown in the lower panel of Fig.~\ref{fig:MeasSpectra}.}
\begin{tabular}{ l  c }
\hline
\hline
Source & Uncertainty [\%] \\
\hline
\hspace{9pt} Trigger efficiency & 1.0 \\
\hspace{9pt} Momentum resolution & 1.0 \\
\hspace{9pt} Particle species composition & 1--10.0 (0.5--5) \\
\hspace{9pt} Track misreconstruction rate & 1.0 \\
\hspace{9pt} Track selection & 1.2--4.0 \\
\hspace{9pt} Spectra relative normalization &  0.0--1.0 \\
\hspace{9pt} Trigger bias & 0.0--4.0 \\
\hline
Total (spectra) & 2.2--10.9 \\
\hline
\hspace{9pt} pp interpolation & 12.0 \\
\hline
Total ($R^{\ast}_\mathrm{pPb}$) & 12.2--16.2 \\
\hline
$\langle \mathrm{T}_\mathrm{pPb} \rangle$  average nuclear thickness & 4.8 \\
\hline
Total ($Y_\text{asym}\quad 0.3< \abs{\etacm} < 0.8 $)& 2.0--3.0 \\
\hline
Total ($Y_\text{asym} \quad 0.8 < \abs{\etacm} < 1.3 $)& 2.0--5.0 \\
\hline
Total ($Y_\text{asym} \quad  1.3 < \abs{\etacm} < 1.8 $) & 2.0--5.0 \\
\hline
\hline
\end{tabular}
\label{tab:systematics}
\end{table}

\section{Results}
\label{sec:results}

\begin{figure}[ht]
  \begin{center}
    \includegraphics[width=\cmsFigWidth]{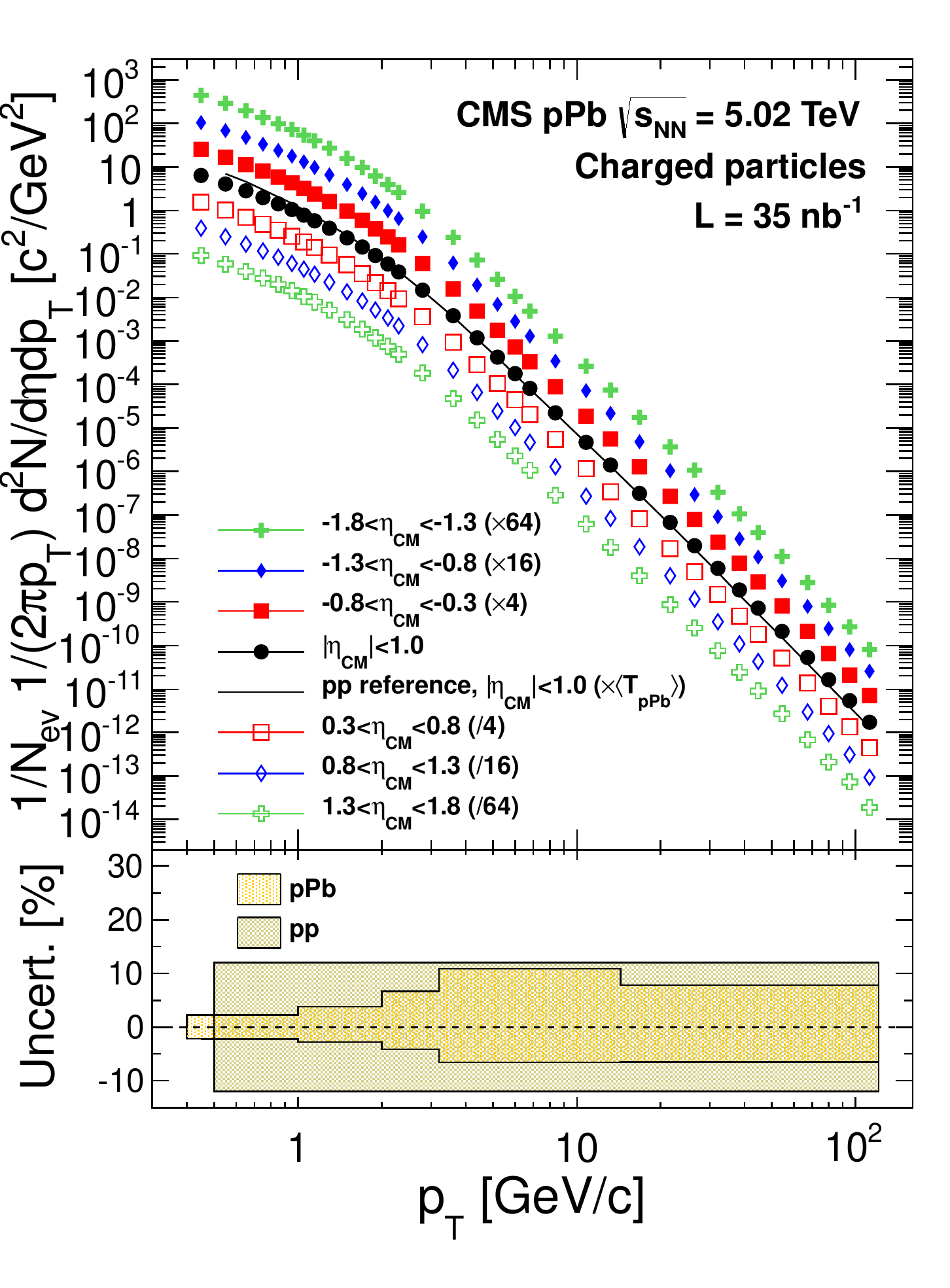}
    \caption{Top: Measured charged-particle transverse momentum spectra in pPb collisions at
$\rtsnn=5.02$\TeV for:
$\abs{\etacm}<1.0$,
$0.3<\pm \etacm<0.8$,
$0.8<\pm \etacm<1.3$,
and $1.3<\pm \etacm<1.8$,
and the interpolated pp reference
spectrum in $\abs{\etacm}<1$, normalized to the number of double-sided events. Positive pseudorapidity
values correspond to the proton beam direction. The spectra have
been scaled by the quoted factors to provide better visibility. Bottom:
Systematic uncertainties in the measured pPb and interpolated pp spectra,
as a function of \pt (see text).}
    \label{fig:MeasSpectra}
  \end{center}
\end{figure}

The measured charged-particle yields in double-sided pPb collisions at $\rtsnn=5.02\TeV$ are plotted
in Fig.~\ref{fig:MeasSpectra} for the
$\abs{\etacm}<1.0$, $0.3<\pm \etacm<0.8$, $0.8<\pm \etacm<1.3$, and $1.3<\pm
\etacm<1.8$ pseudorapidity ranges.
Positive
(negative) pseudorapidity
values correspond to the proton (lead) beam direction.
To improve the visibility of the
results, the spectra at different pseudorapidities have been scaled up and down by multiple factors of 4 relative to the
data for $|\eta_{\rm CM}|<1$. The relative uncertainties for the pPb and the
pp spectra are given in the
bottom panel.

The measurement of the charged-particle nuclear modification factor of Eq.~(\ref{eq:R_AB}) requires
a rescaling of the pp cross section by the average nuclear thickness function
in minimum-bias pPb collisions.
This factor amounts to $\langle \mathrm{T}_\mathrm{pPb}\rangle =(0.0983\pm0.0044)\mbinv$ for inelastic pPb collisions and is obtained
from a Glauber MC simulation~\cite{Miller,PublishedPHOBOSGlauberMC}, where the Pb nucleus is
described using a Woods-Saxon distribution with nuclear radius $6.62\pm0.13$\unit{fm} and skin depth of
$0.546\pm0.055$\unit{fm}~\cite{Miller,DeJager:1987qc}. As double-sided events correspond to 94--97\% of inelastic collisions based on HIJING and EPOS MC computations~\cite{Chatrchyan:2013eya}, the value of $\langle \mathrm{T}_\mathrm{pPb} \rangle$ would be about 5\% higher for double-sided events.

\begin{figure}[ht]
  \begin{center}
    \includegraphics[width=\cmsFigWidth]{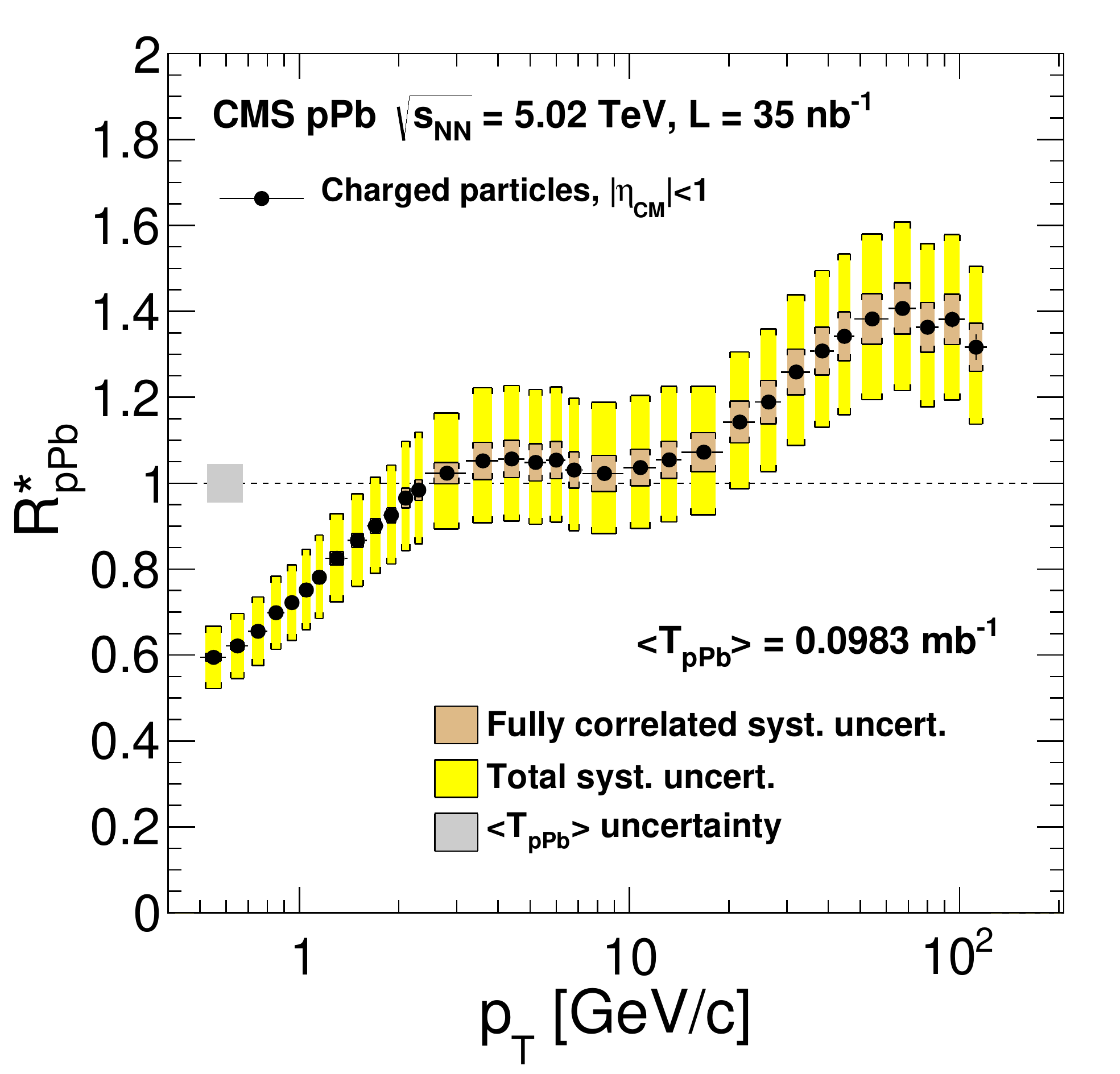}
    \caption{Measured nuclear modification factor as a function of \pt for charged particles
produced in $\abs{\etacm}<1$. The shaded  band at unity and $\pt\approx 0.6$ represents the
uncertainty in the Glauber calculation of $\langle \mathrm{T}_\mathrm{pPb} \rangle$.
The smaller uncertainty band around the data points shows the uncertainty
from effects (combining spectra, track selection, and trigger efficiency) that are fully correlated in
specific \pt regions. The total systematic uncertainties, dominated by uncertainty in the pp
interpolation, are shown by the larger band (see Table~\ref{tab:systematics}). }
   \label{fig:RpPb_CMS}
  \end{center}
\end{figure}

The charged-particle $R^{\ast}_\mathrm{pPb}$ at mid-rapidity ($\abs{\etacm}<1$) is plotted
in Fig.~\ref{fig:RpPb_CMS} as a function of \pt. The shaded  band at unity and $\pt\approx 0.6$
represents the uncertainty in the Glauber calculation of $\langle \mathrm{T}_\mathrm{pPb}\rangle$.
The smaller uncertainty band around the measured values shows the fully correlated uncertainties from the following sources:
spectra relative normalization,
track selection, and trigger efficiency. The total systematic uncertainties are
shown by the larger band around the measured values (Table~\ref{tab:systematics}). The nuclear
modification factor shows a steady rise to unity at $\pt \approx 2\GeVc$, then remains constant at unity up to approximately 20\GeVc, and rises again at higher
\pt, reaching a maximum value around 1.3--1.4 above 40\GeVc.

The fact that the nuclear modification factor is below unity for $\pt\lesssim2$\GeVc is anticipated
since particle production in this region is dominated by softer scattering processes, that are not expected
to scale with the nuclear thickness function.
In the intermediate \pt range (2--5\GeVc),
no significant deviation from unity is found in the $R^{\ast}_\mathrm{pPb}$ ratio.
There are several prior measurements that suggest an interplay of multiple effects in this \pt region.
At lower collision energies, an enhancement (``Cronin effect''~\cite{Cronin:1975})
has been observed~\cite{BRAHMS:2003,PHENIX:2003,PHOBOS:2003,STAR:2003a}
that is larger for baryons than for mesons, and is stronger in the more central collisions. This enhancement has been  attributed to a combination of initial-state multiple scattering effects, causing momentum broadening, and hadronization through parton recombination (a final-state effect)~\cite{Hwa:2009bh} invoked to
accommodate baryon/meson differences.
Recent results from  pPb collisions at $\rtsnn=5.02\TeV$~\cite{CMS:2013,ALICE:2013a,ATLAS:2013,Chatrchyan:2013nka,Chatrchyan:2013eya}
and from dAu collisions at  $\rtsnn$ = 200\GeV ~\cite{Adare:2013esx,Adare:2014keg}
suggest that collective effects may also play a role in the intermediate-\pt region.
Most theoretical models do not predict a Cronin enhancement in
this \pt range at LHC energies as the effect of initial-state multiple scattering
is compensated by nPDF shadowing~\cite{Albacete:2013ei}.

\begin{figure}[ht]
  \begin{center}
    \includegraphics[width=\cmsFigWidth]{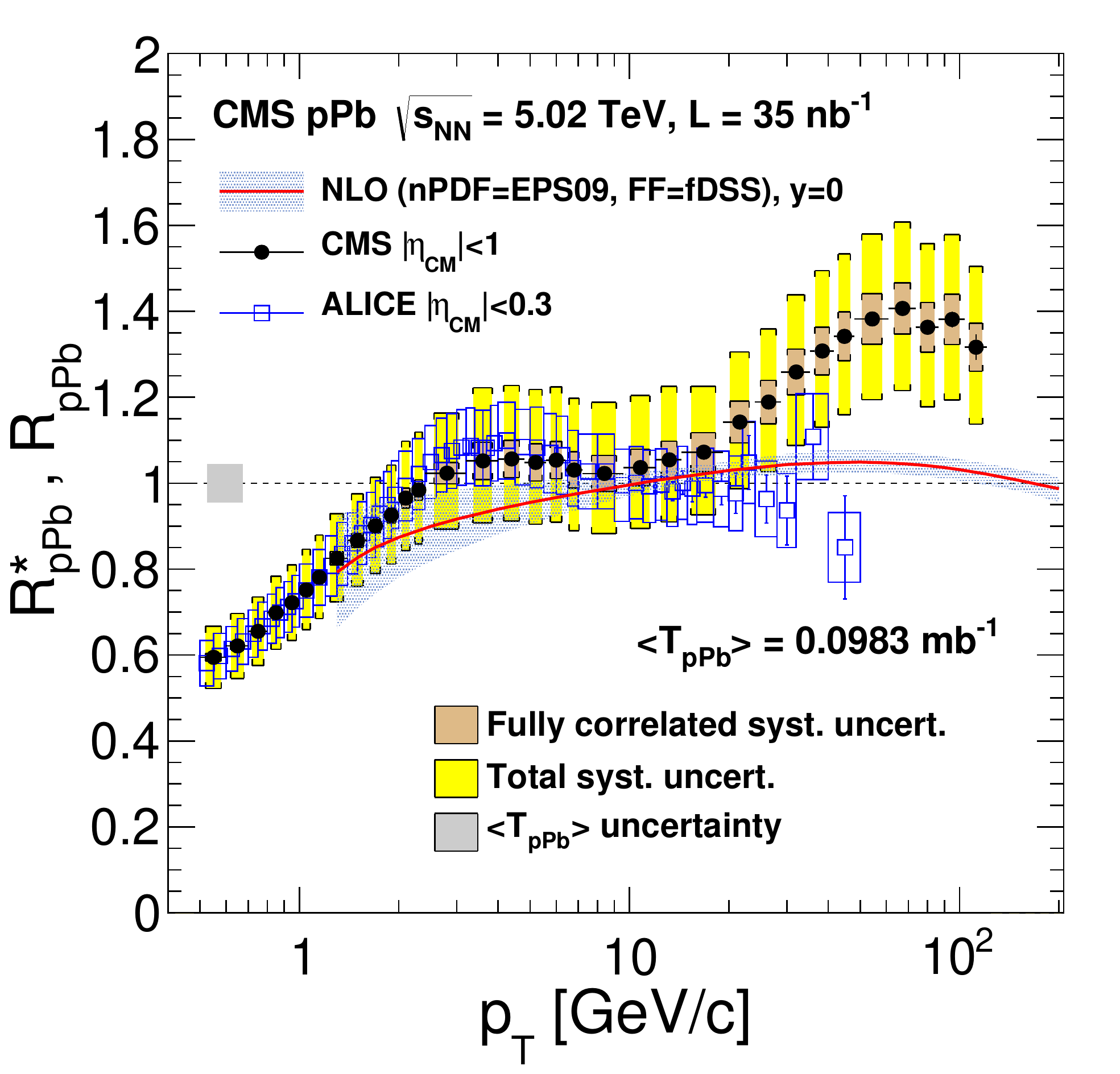}
    \caption{Charged-particle nuclear modification factors measured by CMS in $\abs{\etacm}<1$ (filled circles), and by  ALICE in $\abs{\etacm}<0.3 $ (open squares), are
compared to the NLO pQCD prediction of  Ref.~\cite{Paukkunen:2014vha}.  The theoretical uncertainty is based on the EPS09 error sets. For the CMS measurement, the shaded  band at unity and $\pt\approx 0.6$ represents the
uncertainty in the Glauber calculation of $\langle \mathrm{T}_\mathrm{pPb} \rangle$, the smaller uncertainty band around the data points shows the  fully correlated uncertainties and the total systematic uncertainty is shown by the larger band (see Table~\ref{tab:systematics}). For the ALICE measurement, the total systematic uncertainties, excluding the normalization uncertainty of 6\%, are shown with open boxes.}
    \label{fig:RpPb_CMS_theory_ALICE}
  \end{center}
\end{figure}

In Fig.~\ref{fig:RpPb_CMS_theory_ALICE}, the CMS measurement is compared to the result of an NLO pQCD
calculation~\cite{Paukkunen:2014vha} for charged particles produced at mid-rapidity.
The calculation uses
the CTEQ10~\cite{Lai:2010vv}  free-proton PDF,
the EPS09 nPDF~\cite{Eskola:2009uj},  and
the fDSS fragmentation functions~\cite{deFlorian:2007aj}. The observed rise of
the nuclear modification factor up to $R^{\ast}_\mathrm{pPb} \approx1.3$--1.4 at
high \pt
is stronger than expected theoretically.
None of the available theoretical models~\cite{Albacete:2013ei} predict enhancements beyond $R_\mathrm{pPb} \approx$~1.1
at high \pt. In particular, although the \pt range corresponds to parton momentum fractions
$0.02\lesssim x \lesssim 0.2$, which coincides with the region where parton antishadowing effects are
expected~\cite{Salgado:2011wc},
none of the nPDFs obtained from global fits to nuclear data predict enhancements
beyond 10\% at the large virtualities ($Q^2 \sim \pt^2 \sim 500\text{--}10\,000\GeVcc$) of relevance here.

An
estimate of the significance of this observed
rise above unity for $40 < \pt < 120\GeVc$ is determined by interpreting
all uncertainties as
following a multivariate normal distribution where the components are
the six \pt bins in the kinematic region of interest. The variance of
each component is given by the sum of the statistical and
systematic uncertainties in quadrature. For the case of the
asymmetric
particle species
uncertainty, the smaller negative value is used as the data are uniformly
larger than the expected values of the hypothesis to be tested.
Given that the uncertainties of the reference spectrum are derived from
applying different interpolation procedures and propagating the uncertainties
from previous measurements from multiple experiments, it is not possible
to unambiguously determine how all systematic uncertainties are correlated between
measurements in each \pt bin.
Therefore, a pair of estimates of the possible significance is given. In one case,
 only the systematic uncertainties from
the relative normalization of the spectra, track selection, trigger efficiency,
nuclear thickness function, and NLO pQCD calculation
are treated as fully correlated, while others are treated as uncorrelated.
In the other case, all systematic uncertainties are treated as fully correlated.
Both the hypothesis that $R^{\ast}_\mathrm{pPb}$ is unity and the hypothesis that
$R^{\ast}_\mathrm{pPb}$ is
given by the NLO pQCD calculation are tested.
For the case in which some uncertainties are treated as uncorrelated, a
log-likelihood ratio test is performed using an alternative hypothesis that
$R^{\ast}_\mathrm{pPb}$ is given by either unity or the NLO prediction, scaled by
a constant, \pt-independent, factor.
The hypothesis that $R^{\ast}_\mathrm{pPb}$ is unity
for $40 < \pt < 120\GeVc$ is rejected with a $p$ value of 0.006\%, and the hypothesis
that $R^{\ast}_\mathrm{pPb}$ is given by the NLO pQCD
calculation for $40 < \pt < 120\GeVc$ is rejected with a $p$ value of 0.2\%.
For the case in which all uncertainties are fully correlated,
the log-likelihood ratio test cannot be used, as
the covariance matrix becomes nearly singular and the maximum likelihood estimation
fails.
Instead, a two-tailed univariate
test is performed using the single measurement for $61 < \pt < 74\GeVc$.
From this test, the  hypothesis that $R^{\ast}_\mathrm{pPb}$ is unity for $61 < \pt < 74\GeVc$
is rejected with a
$p$ value of 0.4\%, and the hypothesis that $R^{\ast}_\mathrm{pPb}$ is given by the NLO pQCD
calculation for $61 < \pt < 74\GeVc$ is rejected with a $p$ value of 2\%.

Fig.~\ref{fig:RpPb_CMS_theory_ALICE} also shows the measurement from the ALICE experiment~\cite{Abelev:2014dsa}, which is performed in a narrower pseudorapidity range than the CMS one, and uses a different method (NLO scaling) to obtain the pp reference spectrum based on ALICE pp data measured at $\sqrt{s} = 7$\TeV.  The difference in the
CMS and ALICE $R^{\ast}_\mathrm{pPb}$ results stems primarily from differences in the
charged-hadron spectra measured in pp collisions  at
$\sqrt{s} = 7$\TeV~\cite{QCD_spectrum_09_7,Abelev:2013ala}.

\begin{figure}[ht]
  \begin{center}
    \includegraphics[width=\cmsFigWidth]{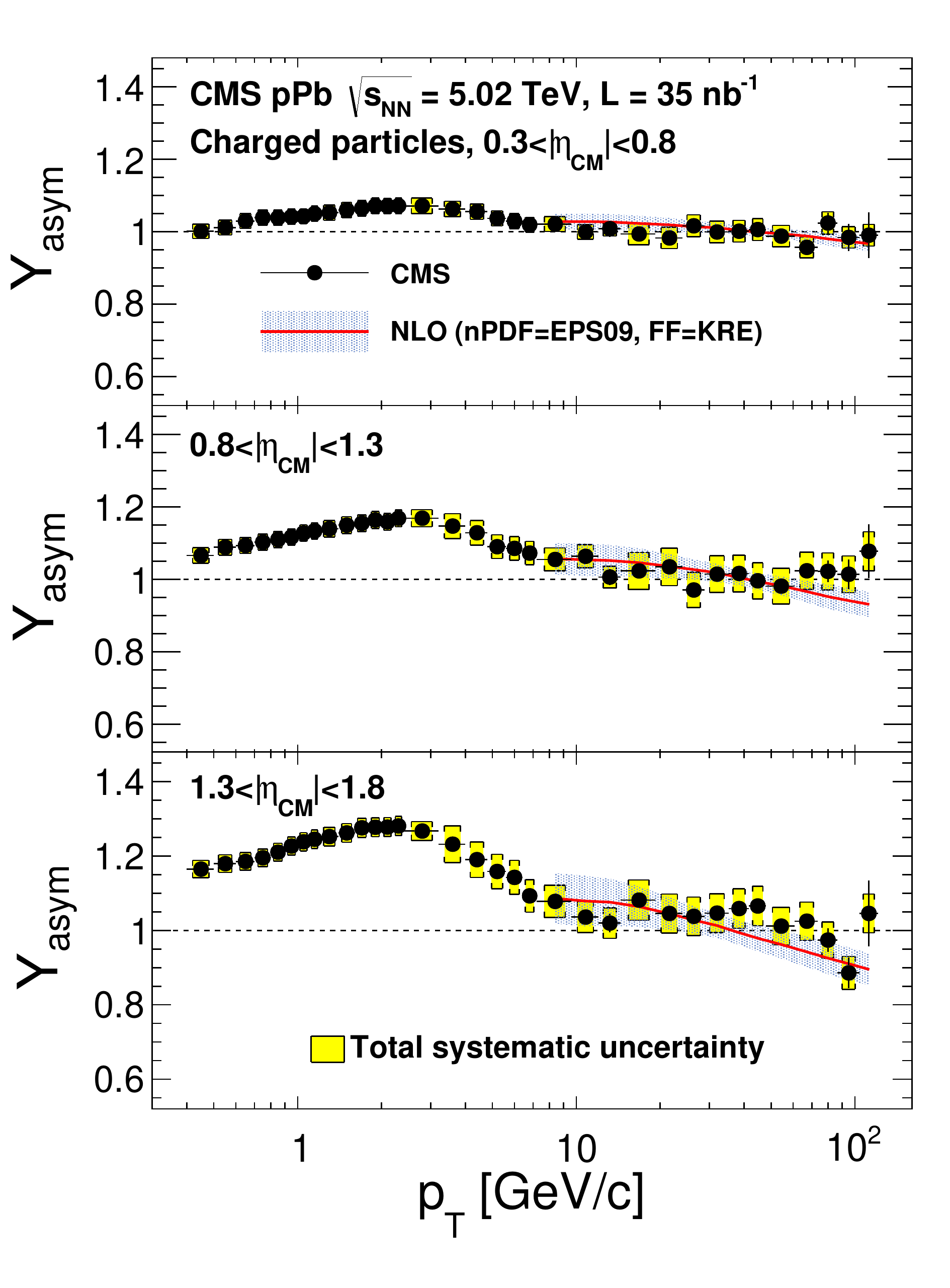}
    \caption{Charged-particle forward-backward yield asymmetry as a function of \pt
for $0.3<\abs{\etacm}<0.8$ (top), $0.8<\abs{\etacm}<1.3$ (middle), and
$1.3<\abs{\etacm}<1.8$ (bottom). The asymmetry is computed as the charged-particle yields in the direction of the Pb beam divided by those of the proton beam.
The solid curves are NLO pQCD theoretical calculations including
nPDFs modifications~\cite{Paukkunen:2014vha}. The theoretical uncertainty is based on the EPS09 error sets.
}
    \label{fig:Y_asym}
  \end{center}
\end{figure}

Figure~\ref{fig:Y_asym} shows the forward-backward yield asymmetry, $Y_\text{asym} $ (Eq.~\ref{eq:Y_asym}), as a function of \pt
for $0.3<\abs{\etacm}<0.8$, $0.8<\abs{\etacm}<1.3$, and
$1.3<\abs{\etacm}<1.8$. In all three $\eta$ ranges, the value of
$Y_\text{asym}$ rises from $\pt\approx0.4$ to about 3\GeVc, then falls to unity
at a \pt of 10\GeVc, and remains constant at unity up to the highest \pt values.
At the lowest
\pt value, $Y_\text{asym}$ is consistent with unity for
$0.3<\abs{\etacm}<0.8$, but is above unity in the larger pseudorapidity
regions. For $\pt < 10 \GeVc$, the $Y_\text{asym}$ is larger than unity as has been predicted by models including nuclear shadowing~\cite{Albacete:2013ei}.
A theoretical NLO pQCD computation
of $Y_\text{asym}$ at high \pt~\cite{Paukkunen:2014vha},
using CTEQ6~\cite{Nadolsky:2008zw} free-proton PDFs,
EPS09 nPDFs~\cite{Eskola:2009uj}, and Kretzer parton-to-hadron fragmentation functions~\cite{Kretzer:2000yf},
is also shown in Fig.~\ref{fig:Y_asym}.
 The theoretical predictions are consistent with these data.

To determine if the $R^{\ast}_\mathrm{pPb}$ and $Y_\text{asym}$ results
can be consistently interpreted in terms of nPDF modifications,
an MC study using the \PYTHIA (Z2 tune) event generator was performed to
correlate each high-\pt hadron to the fractional momentum, $x$, of the initial-state parton from the Pb nucleus that participated in the hard-scattering process
producing the final hadron.
In all pseudorapidity intervals studied here, most of the hadrons with
 $\pt\gtrsim 20$\GeVc, \ie, in the range where the
$R^{\ast}_\mathrm{pPb}$ exceeds unity in Fig.~\ref{fig:RpPb_CMS}, come from the $x$ region that is
associated with antishadowing in the nPDF distributions. Although the mean of the
$x$ distribution increases with $\etacm$, for hadrons
with \pt above 20\GeVc it remains in the
range $0.02\lesssim x \lesssim 0.2$. Thus, similar antishadowing effects are expected in the
positive and negative $\etacm$ regions resulting in a $Y_\text{asym}$
close to unity.
At low \pt, corresponding to $x \lesssim 0.02$, a larger hadron yield
is observed in the direction of the Pb beam. This is qualitatively
consistent with expectations of gluon shadowing~\cite{Albacete:2013ei}.

An enhancement in $R^{\ast}_\mathrm{pPb}$ at high \pt can possibly arise if the
quark-jet fraction is larger in pPb than in pp collisions.
Since the charged-particle products of quark fragmentation more often
have higher relative \pt than those produced by gluon fragmentation,
that could lead to an enhancement in the charged-particle production at
high \pt
beyond NLO expectations, without a corresponding increase in the jet
$R_\mathrm{pPb}$~\cite{Aad:2014bxa,Adam:2015hoa}.
We note that the gluon-to-hadron fragmentation functions are not well
constrained
in pp collisions at LHC energies~\cite{d'Enterria:2013vba}, although such
uncertainties should largely cancel in ratios of cross sections.

\section{Summary}
\label{sec:summary}

Charged-particle spectra have been measured in pPb collisions at
$\rtsnn=5.02$\TeV in the transverse momentum range of $
0.4 < \pt < 120$\GeVc for pseudorapidities up to
$\abs{\etacm} = 1.8$. The forward-backward
yield asymmetry has been measured as a function of \pt for three bins in
$\etacm$. At $\pt< 10$ \GeVc, the charged-particle
production is enhanced in the direction of the Pb beam, in qualitative
agreement with nuclear shadowing expectations. The nuclear modification factor at
mid-rapidity, relative to a reference spectrum interpolated from pp
measurements at lower and higher collision energies, rises above unity
at high \pt reaching an $R^{\ast}_\mathrm{pPb}$ value of 1.3--1.4 at
$\pt\gtrsim40$\GeVc.
The observed enhancement is larger than expected
from NLO pQCD predictions that include antishadowing effects in the
nuclear PDFs in this kinematic range. Future direct measurement of the spectra of jets and charged particles in pp collisions at a center-of-mass energy of 5.02\TeV
is necessary to better constrain the fragmentation functions and also to
reduce the dominant systematic uncertainties in the charged-particle nuclear modification factor.

\begin{acknowledgments}
\hyphenation{Bundes-ministerium Forschungs-gemeinschaft Forschungs-zentren} We congratulate our colleagues in the CERN accelerator departments for the excellent performance of the LHC and thank the technical and administrative staffs at CERN and at other CMS institutes for their contributions to the success of the CMS effort. In addition, we gratefully acknowledge the computing centres and personnel of the Worldwide LHC Computing Grid for delivering so effectively the computing infrastructure essential to our analyses. Finally, we acknowledge the enduring support for the construction and operation of the LHC and the CMS detector provided by the following funding agencies: the Austrian Federal Ministry of Science, Research and Economy and the Austrian Science Fund; the Belgian Fonds de la Recherche Scientifique, and Fonds voor Wetenschappelijk Onderzoek; the Brazilian Funding Agencies (CNPq, CAPES, FAPERJ, and FAPESP); the Bulgarian Ministry of Education and Science; CERN; the Chinese Academy of Sciences, Ministry of Science and Technology, and National Natural Science Foundation of China; the Colombian Funding Agency (COLCIENCIAS); the Croatian Ministry of Science, Education and Sport, and the Croatian Science Foundation; the Research Promotion Foundation, Cyprus; the Ministry of Education and Research, Estonian Research Council via IUT23-4 and IUT23-6 and European Regional Development Fund, Estonia; the Academy of Finland, Finnish Ministry of Education and Culture, and Helsinki Institute of Physics; the Institut National de Physique Nucl\'eaire et de Physique des Particules~/~CNRS, and Commissariat \`a l'\'Energie Atomique et aux \'Energies Alternatives~/~CEA, France; the Bundesministerium f\"ur Bildung und Forschung, Deutsche Forschungsgemeinschaft, and Helmholtz-Gemeinschaft Deutscher Forschungszentren, Germany; the General Secretariat for Research and Technology, Greece; the National Scientific Research Foundation, and National Innovation Office, Hungary; the Department of Atomic Energy and the Department of Science and Technology, India; the Institute for Studies in Theoretical Physics and Mathematics, Iran; the Science Foundation, Ireland; the Istituto Nazionale di Fisica Nucleare, Italy; the Ministry of Science, ICT and Future Planning, and National Research Foundation (NRF), Republic of Korea; the Lithuanian Academy of Sciences; the Ministry of Education, and University of Malaya (Malaysia); the Mexican Funding Agencies (CINVESTAV, CONACYT, SEP, and UASLP-FAI); the Ministry of Business, Innovation and Employment, New Zealand; the Pakistan Atomic Energy Commission; the Ministry of Science and Higher Education and the National Science Centre, Poland; the Funda\c{c}\~ao para a Ci\^encia e a Tecnologia, Portugal; JINR, Dubna; the Ministry of Education and Science of the Russian Federation, the Federal Agency of Atomic Energy of the Russian Federation, Russian Academy of Sciences, and the Russian Foundation for Basic Research; the Ministry of Education, Science and Technological Development of Serbia; the Secretar\'{\i}a de Estado de Investigaci\'on, Desarrollo e Innovaci\'on and Programa Consolider-Ingenio 2010, Spain; the Swiss Funding Agencies (ETH Board, ETH Zurich, PSI, SNF, UniZH, Canton Zurich, and SER); the Ministry of Science and Technology, Taipei; the Thailand Center of Excellence in Physics, the Institute for the Promotion of Teaching Science and Technology of Thailand, Special Task Force for Activating Research and the National Science and Technology Development Agency of Thailand; the Scientific and Technical Research Council of Turkey, and Turkish Atomic Energy Authority; the National Academy of Sciences of Ukraine, and State Fund for Fundamental Researches, Ukraine; the Science and Technology Facilities Council, UK; the US Department of Energy, and the US National Science Foundation.
Individuals have received support from the Marie-Curie programme and the European Research Council and EPLANET (European Union); the Leventis Foundation; the A. P. Sloan Foundation; the Alexander von Humboldt Foundation; the Belgian Federal Science Policy Office; the Fonds pour la Formation \`a la Recherche dans l'Industrie et dans l'Agriculture (FRIA-Belgium); the Agentschap voor Innovatie door Wetenschap en Technologie (IWT-Belgium); the Ministry of Education, Youth and Sports (MEYS) of the Czech Republic; the Council of Science and Industrial Research, India; the HOMING PLUS programme of Foundation for Polish Science, cofinanced from European Union, Regional Development Fund; the Compagnia di San Paolo (Torino); the Consorzio per la Fisica (Trieste); MIUR project 20108T4XTM (Italy); the Thalis and Aristeia programmes cofinanced by EU-ESF and the Greek NSRF; and the National Priorities Research Program by Qatar National Research Fund.
\end{acknowledgments}

\bibliography{auto_generated}

\cleardoublepage \appendix\section{The CMS Collaboration \label{app:collab}}\begin{sloppypar}\hyphenpenalty=5000\widowpenalty=500\clubpenalty=5000\textbf{Yerevan Physics Institute,  Yerevan,  Armenia}\\*[0pt]
V.~Khachatryan, A.M.~Sirunyan, A.~Tumasyan
\vskip\cmsinstskip
\textbf{Institut f\"{u}r Hochenergiephysik der OeAW,  Wien,  Austria}\\*[0pt]
W.~Adam, T.~Bergauer, M.~Dragicevic, J.~Er\"{o}, M.~Friedl, R.~Fr\"{u}hwirth\cmsAuthorMark{1}, V.M.~Ghete, C.~Hartl, N.~H\"{o}rmann, J.~Hrubec, M.~Jeitler\cmsAuthorMark{1}, W.~Kiesenhofer, V.~Kn\"{u}nz, M.~Krammer\cmsAuthorMark{1}, I.~Kr\"{a}tschmer, D.~Liko, I.~Mikulec, D.~Rabady\cmsAuthorMark{2}, B.~Rahbaran, H.~Rohringer, R.~Sch\"{o}fbeck, J.~Strauss, W.~Treberer-Treberspurg, W.~Waltenberger, C.-E.~Wulz\cmsAuthorMark{1}
% ...
\vskip\cmsinstskip
% ...
1:~~Also at Vienna University of Technology, Vienna, Austria\\
2:~~Also at CERN, European Organization for Nuclear Research, Geneva, Switzerland\\
% ...

\end{sloppypar}
\end{document}